\documentclass[runningheads]{llncs}
\usepackage{graphicx}
\usepackage{amsmath}
\usepackage{amssymb}
\usepackage{bm}
\usepackage{hyperref} 
\newcommand*\samethanks[1][\value{footnote}]{\footnotemark[#1]}

\begin{document}
\title{Self-Attentive Adversarial Stain Normalization}
%
%

\author{Aman Shrivastava\inst{1} \and
William Adorno\inst{1} \and
Yash Sharma\inst{1} \and
Lubaina Ehsan\inst{1} \and \\
S. Asad Ali\inst{2} \and
Sean R. Moore\inst{1} \and
Beatrice Amadi\inst{3} \and
Paul Kelly\inst{3, 4} \and
Sana Syed \inst{1} \thanks{Co-corresponding authors} \and
Donald E. Brown \inst{1} \samethanks}
\authorrunning{Shrivastava et al.}
%
\institute{University of Virginia, Charlottesville, Virginia, USA \and
Aga Khan University, Karachi, Pakistan \and
University of Zambia School of Medicine, Lusaka, Zambia \and
Queen Mary University of London, London, England\\
\email{\{as3ek, wa3mr, ys5hd\}@virginia.edu}}
\maketitle              
\begin{abstract}
Hematoxylin and Eosin (H\&E) stained Whole Slide Images (WSIs) are utilized for biopsy visualization-based diagnostic and prognostic assessment of diseases. Variation in the H\&E staining process across different lab sites can lead to important variations in biopsy image appearance. These variations introduce an undesirable bias when the slides are examined by pathologists or used for training deep learning models. Traditionally proposed stain normalization and color augmentation strategies can handle the human level bias. But deep learning models can easily disentangle the linear transformation used in these approaches, resulting in undesirable bias and lack of generalization. To handle these limitations, we propose a Self-Attentive Adversarial Stain Normalization (SAASN) approach for the normalization of multiple stain appearances to a common domain. This unsupervised generative adversarial approach includes self-attention mechanism for synthesizing images with finer detail while preserving the structural consistency of the biopsy features during translation. SAASN demonstrates consistent and superior performance compared to other popular stain normalization techniques on H\&E stained duodenal biopsy image data.

\end{abstract}
\section{Introduction}
Histopathology involves staining patient biopsies for microscopic inspection to identify visual evidence of diseases. The most widely used stain in histopathology is the Hematoxylin and Eosin (H\&E) stain \cite{fischer2008hematoxylin}. Hematoxylin has a deep blue-purple color and stains acidic structures such as nucleic acids (DNA in cell nuclei). While Eosin is red-pink, and stains basic structures such as nonspecific proteins in the cytoplasm and the stromal matrix. Staining is crucial as it enables visualization of the microscopic structural features in the biopsy. The process of staining is followed by glass biopsy slide creation and eventually digitization into Whole Slide Images (WSIs) using digital scanners.

Computer vision is becoming increasingly useful in the field of histology for computed-aided diagnosis and discovering information about histopathological microscopic cellular \cite{litjens2017survey}. Tremendous potential has been shown for training deep learning algorithms on these datasets for diagnosis and visual understanding of diseases requiring histopathological assessment. Convolution Neural Networks (CNNs) have been successfully reported for biopsy-based diagnosis of breast cancer and enteropathies among others \cite{45922,wei2019automated}. The performance and fairness of such data-driven methods is dependent on the data used for training. Therefore, it is imperative for the training data to be free of any bias that might skew the models. A common source of such bias is significant stain color variation among images. This is due to the discrepancies in the manufacturing protocol and the raw materials of the staining chemicals \cite{bejnordi2014quantitative} across different sites where the biopsy slides are prepared. Multiple H\&E stain distributions within the CNN input data can lead to biased predictions where the results are influenced by color differences rather than microscopic cellular features of interest for clinical diagnostic interpretation. Additionally, it causes difficulty for a trained model to make predictions on a biopsy WSI with a new stain appearance that is not represented in the data used to train the model.

To overcome these issues, researchers have developed stain normalization techniques to convert all input images to an equivalent color distribution. Some of the most popular stain normalization techniques depend on a qualitatively chosen target image that represents an ideal color appearance \cite{macenko2009method,khan2014nonlinear,vahadane2016structure}. The input (source) image is normalized to match the stain profile of the chosen target image. The obvious downside to this approach is that the normalization is highly dependent on the color distribution of a single image. Rather than using just one target image to represent an entire stain distribution, an alternative approach to consider an entire set of images that share the same stain distribution as the target domain has been suggested \cite{janowczyk2017stain,shaban2019staingan}. A mapping function can then be learned to translate images from a particular source domain to a target domain. This problem can be modelled as an unsupervised image-to-image translation task \cite{liu2017unsupervised}. 

Recently, Generative Adversarial Networks (GANs) have been shown to demonstrate exceptional results in unpaired image translation tasks \cite{yi2017dualgan,cyclegan,kim2017learning}. However, the challenge posed by the stain normalization task is to ensure the preservation of fine details and microscopic structural properties that are crucial for the correct disease assessment. Additionally, since the biopsy slides can be sourced from multiple sites, the framework needs to be capable of mapping multiple stain distributions to a common target distribution.

In this paper, we propose a novel adversarial approach that can execute \textit{many-to-one} domain stain normalization. A custom loss function, structural cycle-consistency loss, is designed to make sure that the structure of the image is preserved during translation. Self-attention \cite{parikh2016decomposable} is used to ensure that highly detailed microscopic features can be synthesized in the image. Our approach and other leading stain normalization techniques are compared on duodenum biopsy image data that was used to diagnose Celiac or Environmental Enteropathy disease in children, and on MITOS-ATYPIA Challenge dataset consisting of H\&E stained WSI slides scanned by two scanners: Aperio Scanscope XT and Hamamatsu Nanozoomer 2.0-HT. SAASN demonstrated superior performance in preserving the structural integrity of images while transferring the stain distribution from one domain to the other.


\section{Related Work}

The earliest methods that attempted stain normalization were primarily simple style transfer techniques. Histogram specification mapped the histogram statistics of the target image with the histogram statistics of the source \cite{coltuc2006exact}. This approach only works well if the target and source images have similar color distributions. Forcing the normalization of the source image to match the histogram statistics of the target can create artifacts which can alter the structural integrity. As demonstrated by Reinhard \cite{reinhard2001color}, color transfer with histogram specification can also be performed in a decorrelated CIELAB color space which is designed to approximate the human visual system.

For H\&E stained histology images, the presence of each stain or the lack thereof at each pixel should represent the most appropriate color space. Considering this, researchers developed stain normalization methods that outperformed the histogram specification technique by leveraging stain separation. These techniques start with converting an RGB image into Optical Density ($O_D$) as $O_D = log\frac{I_0}{I}$, where $I_0$ is the total possible illumination intensity of the image and $I$ is the RGB image. Color Deconvolution (CD) is made easier in the OD space, because the stains now have a linear relationship with the OD values. The CD is typically expressed as $O_D = VS$, where $V$ is the matrix of stain vectors and $S$ is the stain density map. The stain density map can preserve the cell structures of the source image, while the stain vectors are updated to reflect the stain colors of the target image. 

In Macenko \cite{macenko2009method}, stain separation is computed using singular value decomposition on the OD tuples. Planes are created from the two largest singular values to represent H\&E stains. One useful assumption with this approach is that the color appearance matrix is non-negative, this is the case because a stain value of zero would refer to the stain not being present at all. The approach by Vahadane \cite{vahadane2016structure} also includes the non-negative assumption, as well as, a sparsity assumptions, which states that each pixel is characterized by an effective stain that relates to a particular cell structure (nuclei cells, cytoplasm, etc.). Stain separation is generated with Sparse Non-negative Matrix Factorization (SNMF) where the sparsity acts as a constraint to greatly reduce the solution space \cite{roy2018study}. SNMF is calculated using dictionary learning via the SPAMS package.

While Macenko and Vahadane are both unsupervised techniques, supervised approaches to this problem have also been studied. Khan \cite{khan2014nonlinear} applies a relevance vector machine or a random forest model to classify each pixel as hematoxylin, eosin or background. The authors provide a pre-trained model for cases which is only useful if the color distribution of new source images is close to the color distribution of their training data. Training a new model would require a training set with pixel level annotations for each stain. After the stain separation, the color of the target image is mapped with a non-linear spline. The non-linear mapping approach can lead to undesirable artifacts and this normalization approach is more computationally costly than the unsupervised approaches. 

Recently, techniques for stain normalization have progressed to include non-linear approaches \cite{shaban2019staingan,bentaieb2017adversarial,janowczyk2017stain,khan2014nonlinear,zanjani2018stain}. The StainGAN \cite{shaban2019staingan} approach applied the CycleGAN framework for \textit{one-to-one} domain stain transfers. In a \textit{one-to-one} stain transfer situation, the cycle-consistency loss is calculated by taking the $L1$ distance between the cycled image and the ground truth. In a \textit{many-to-one} situation, the cycled image will likely have a different color appearance than the original image. Therefore, a new loss function that focuses on image structure and not the color differences is required.

Biopsy WSIs contain repetitive patterns across the image in the form of recurring cell structures, stain gradients, and background alike. During translation, these spatial dependencies can be used to synthesize realistic images with finer details. Self-attention \cite{parikh2016decomposable} exhibits impressive capability in modelling long-range dependencies in images. SAGAN \cite{sagan} demonstrated the use of self-attention mechanism into convolutional GANs to synthesize images in a class conditional image generation task. We incorporate these advances in SAASN to enable it to efficiently find spatial dependencies in different areas of the image.

\section{Approach}

\begin{figure}
\centering
\includegraphics[width=0.9\textwidth]{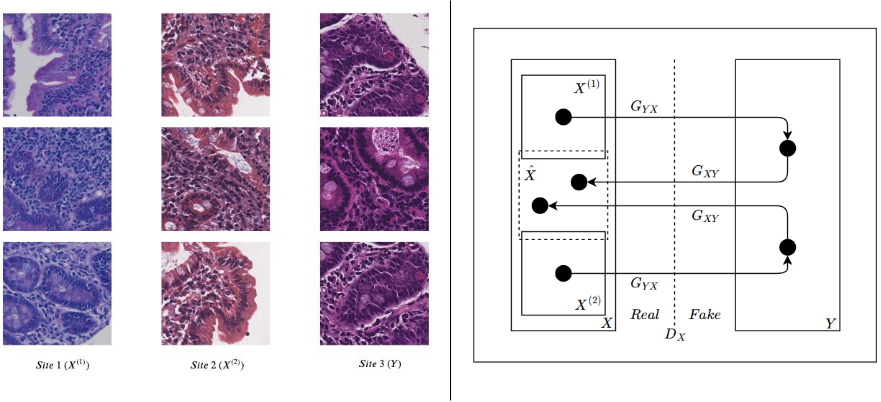}
\caption{\textbf{(Left)} H\&E stained duodenal biopsy patches created from whole slide images sourced from different locations. \textbf{(Right)} Visual example of a \textit{many-to-one} stain transfer network. Two different stains are present as inputs within $X$: $X^{(1)}$ and $X^{(2)}$. Both of these domains are translated to $Y$ with $G_{XY}$. To complete the cycle, $G_{YX}$ returns the image back to the $X$ domain, but it can no longer be mapped directly to the input sub-domains $X^{(1)}$ or $X^{(2)}$ from which it originated. Instead, the image is mapped back to $\hat{X}$ which is represents a new domain of stain appearance.} \label{fig1}
\end{figure}

The general objective of the proposed framework is to learn the mapping between stain distributions represented by domains $X$ and $Y$. Since the aim of the approach is to normalize stain patterns across the entire dataset, one of these domains can be considered as the target domain (say $Y$). The task is then to generate images that are indistinguishable the target domain images based on stain differences. The stain normalization task desires translation of images to a singular domain of stain distribution. This allows us to have multiple sub-domains in domain $X$ representing different stain patterns. The overall objective then becomes to learn mapping functions $G_{YX}: X \rightarrow Y$ and $G_{XY}: Y \rightarrow X$ given unpaired training samples $\{x_i^k\}_{i=1}^N$, $x_i^{k} \in X^{(k)} \in X$, $k \in \left [ 1, K \right ]$ where $K$ denotes the number of sub-domains in $X$ and $\{y_j\}_{j=1}^M$, $y_j \in Y$. The distribution of the training dataset is denoted as $x \sim p\left(x \mid k \right)$ and $y \sim p\left(y \right)$. Additionally, two discriminator functions $D_X$ and $D_Y$ are used. $D_X$ is employed to distinguish mapped images $G_{XY} \left( y_i \right)$ from $x_i$ while in a similar fashion $D_Y$ is used to distinguish $G_{YX} \left( x_i \right)$ from $y_i$. As illustrated in Fig.~\ref{fig1}, the mapping function $G_{XY}$ will map images from domain $Y$ to a previously undefined sub-domain $\hat{X}$ whose boundary is defined by the optimization function and the training data distributions in domain $X$. The overall optimization function used to train the designed framework includes a combination of \textit{adversarial loss} \cite{gan}, \textit{cycle consistency loss} \cite{cyclegan}, \textit{identity loss} \cite{identity}, \textit{structural cycle consistency loss} based on the \textit{structural similarity index} \cite{ssim} and a \textit{discriminator boundary control} factor. 

\textbf{Adversarial loss} is used to ensure that the stain distribution of the generated images matches the distribution of the real (ground truth) images in that domain. The objective for the mapping function $G_{YX} : X \rightarrow Y$ and the corresponding discriminator $D_Y$ is defined as: 
\begin{equation}
    \mathcal{L}_{adv}^{Y} = \mathbb{E}_{y \sim p_{\left(y \right)}}[\log D_Y(y)] 
    + \mathbb{E}_{x \sim p_{\left(x \mid k \right)}}[\log \left ( 1 - D_Y(G_{YX}(x)) \right )]
\end{equation}
Here $G_{YX}$ tries to generate images that are indistinguishable from images in domain $Y$ and consequently fool the discriminator $D_Y$, i.e. the generator $G_{YX}$ tries to minimize the given objective function while the discriminator $D_Y$ tries to maximize it. Similarly the objective for the reverse mapping function $G_{XY} : Y \rightarrow X$ is defined. The presence of multiple distinct stain distributions in the domain $X$ can make it challenging for the discriminator $D_X$ to learn the decision boundary surrounding the domain $X$. This can especially pose a challenge when there is an overlap or proximity in the stain distribution of one of the sub-domains of $X$ and the target domain $Y$ in the high-dimensional space. Therefore, to make sure that the decision boundary learned by $D_X$ does not include sections of the target domain $Y$, a \textbf{discriminator boundary control} factor is added to the optimization function as follows:
\begin{multline}
    \mathcal{L}_{adv}^{X} = \mathbb{E}_{x \sim p_{\left(x \mid k \right)}}[\log D_X(x)]
    + \mathbb{E}_{y \sim p_{\left(y \right)}}[\log \left ( 1 - D_X(G_{XY}(y)) \right )] \\
    + \mathbb{E}_{y \sim p_{\left(y \right)}}[\log \left ( 1 - D_X(y) \right )]
\end{multline}

\textbf{Cycle consistency loss} \cite{cyclegan} is implemented to reconcile with the unpaired nature of the task. To overcome the lack of a ground truth image for a fake image generated in a particular domain, the image is mapped back to its original domain using the reverse mapping function. The reconstructed image is then compared to the original source image to optimize the mapping function as follows: 
\begin{equation}
    \mathcal{L}_{cyc} = \mathbb{E}_{x \sim p\left( x \mid k\right)} \left[ \left \| G_{XY}(G_{YX}(x)) - x \right \|_1 \right]
    + \mathbb{E}_{y \sim p\left( y\right)} \left[ \left \| G_{YX}(G_{XY}(y)) - y \right \|_1 \right]
\end{equation}

\textbf{Structural cycle consistency loss} is added to the objective function to alleviate the shortcomings of the cycle consistency loss for \textit{many-to-one} translation. In a \textit{many-to-one} situation, the cycled images are likely to have a distinct color distribution than any of the sub-domains. Therefore minimizing the $L1$ distance between original and the cycled image alone is not an effective way to ensure cycle consistency. We use a color agnostic structural dissimilarity loss based on the Structural Similarity (SSIM) index \cite{ssim} as follows: 

\begin{equation}
    \mathcal{L}_{scyc} = \frac{1 - SSIM \left( G_{XY}(G_{YX} \left( x \right)), x \right)}{2}
    + \frac{1 - SSIM \left( G_{YX}(G_{XY} \left( y \right)), y \right)}{2}
\end{equation}

Additionally, to ensure that the the mapping learnt by the generator does not result in the loss of biological artifacts, the structural dissimilarity loss is also computed between the mapped and the original image: 
\begin{equation}
    \mathcal{L}_{dssim} = \frac{\left(1 - SSIM \left( G_{YX} \left( x \right), x \right) \right)}{2}
    + \frac{\left(1 - SSIM \left( G_{XY} \left( y \right), y \right) \right)}{2}
\end{equation}
where
\begin{equation}
  SSIM(a,b) = \frac{(2\mu_a\mu_b + C_1) + (2 \sigma _{ab} + C_2)} 
    {(\mu_a^2 + \mu_b^2+C_1) (\sigma_a^2 + \sigma_b^2+C_2)}
\end{equation}
where $\mu$, $\sigma$ are the respective means and standard deviations of the windows ($a$ and $b$) of the fixed size $N\times N$ that strides over the input image. $C_1$ and $C_2$ are stabilizing factors that prevent the denominator from disappearing. These measures are calculated for multiple corresponding windows of gray-scaled input images and aggregated to get the final measure. Gray-scaled inputs are used to focus on structural differences between images and not changes in color.

\textbf{Identity loss} \cite{identity} is utilized to regularize the generator and preserve the overall composition of the image. The generators are rewarded if a near identity mapping is produced when an image from the respective target domain is provided as an input image. In other words, when an image is fed into a generator of its own domain, the generator should produce an image that is nearly identical to the input. This is enforced by minimizing the $L1$ distance of the resulting image with the input image as follows:
\begin{equation}
    \mathcal{L}_{id} = \mathbb{E}_{y \sim p(y)}\left[ \left \| G_{YX}(y) - y \right \|_1\right]
    + \mathbb{E}_{x \sim p(x \mid k)}\left[ \left \| G_{XY}(x) - x \right \|_1\right]
\end{equation}

The overall objective function then becomes:
\begin{equation}
    \mathcal{L}(G_{YX}, G_{XY}, D_{X}, D_{Y}) = \mathcal{L}_{adv}^{Y} + \mathcal{L}_{adv}^{X}
    + \alpha * \mathcal{L}_{cyc} + \beta * \mathcal{L}_{scyc} + \gamma * \mathcal{L}_{dssim} + \delta * \mathcal{L}_{id}  
\end{equation}
where parameters $\alpha$, $\beta$, $\gamma$ and $\delta$ manage the importance of different loss terms. The parameters in the generators and the discriminators are tuned by solving the above objective as:
\begin{equation}
    G_{YX}^* , G_{XY}^* = 
    \arg \min_{G_{YX}, G_{XY}} \max_{D_{X}, D_{Y}} \mathcal{L}(G_{YX}, G_{XY}, D_{X}, D_{Y})
\end{equation}
In the following sections, we describe the implementation and compare our results with other current state-of-the-art methods of color normalization with both multiple ($K = 2$) and single ($K = 1$) sub-domains in $X$. 


\begin{figure*}[t]
\centering
\includegraphics[width=\textwidth]{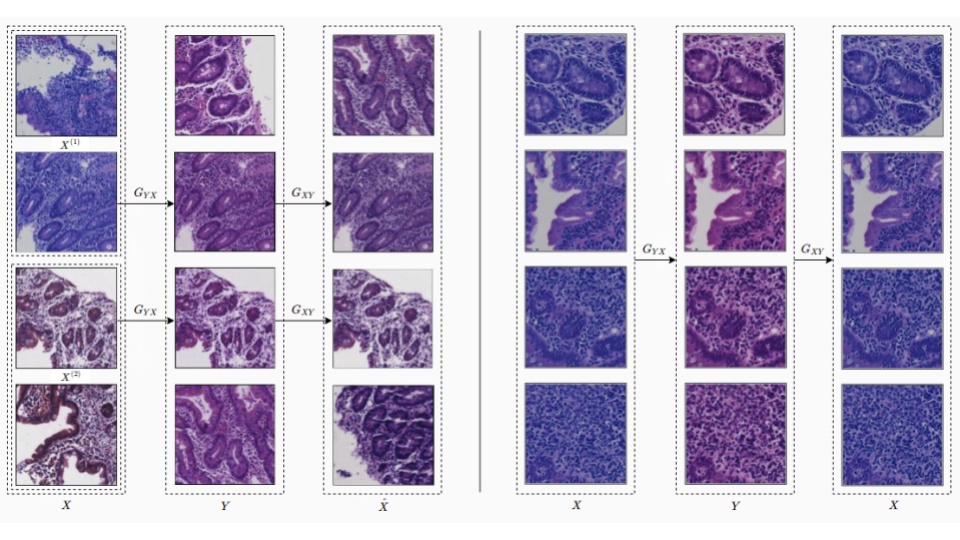} 
\caption{\textit{Left}: Results when mapping was done from two sub-domains of $X$ to $Y$. Patches from both domains $X^{(1)}$ and $X^{(2)}$ are translated to domain $Y$ using $G_{YX}$. These generated images are then translated back to a new domain defined by a $G_{XY}$ as a combination of stain distributions of sub-domains of $X$. Patches on either end of the second column are real images from domain $Y$ and have been added to visually show the performance of $G_{YX}$. \textit{Right}: Results when mapping was learnt using a single domain in $X$ to $Y$. }
\label{fig3}
\end{figure*}

\section{Dataset and Implementation}

\subsection{Dataset}
For this paper, the algorithm was evaluated on two datasets: 1) duodenal biopsy patches were extracted from 465 high resolution WSIs from 150 H\&E stained duodenal biopsy slides (where each glass slide could have one or more biopsies). The biopsies were from patients with Celiac Disease (CD) and Environmental Enteropathy (EE). The biopsies were from children who underwent endoscopy procedures at either Site 1 (Aga Khan University, Pakistan, 10 children, n = 34 WSIs), Site 2 (University Teaching Hospital, Zambia, 16 children, n = 19 WSIs), or Site 3 (University of Virginia Childrens Hospital, 63 children, n = 236 WSIs; and 61 healthy children, n = 173 WSI). It was observed that there was a significantly large stain variation between images originating from different sites. While images from Site 1 were different tones of dark blue, images from Site 3 were more pink with images from Site 2 lying somewhere in the middle of this spectrum. Our approach and other competing methods were performed on $500\times 500$ pixel patches generated from the images, which were further resized to $256\times 256$ pixel to marginally reduce the resolution. In the multi-sub-domain setup, patches from Site 1 (sub-domain $X^{(1)}$) and Site 2 (sub-domain $X^{(2)}$) were both considered to be in domain $X$ and patches from Site 3 to be in domain $Y$. While in single sub-domain training setup, patches from Site 1 were considered to be in domain $X$ and Site 3 to be in domain $Y$. For training both $X$ and $Y$ had $16000$ patches where $X^{(1)}$ contributed $10817$ and $X^{(2)}$ $5183$ patches. Testing metrics were computed on $1500$ patches in each sub-domain.

2) The algorithm was also evaluated on a publicly available MITOS-ATYPIA 14 challenge dataset\footnote{https://mitos-atypia-14.grand-challenge.org/} to demonstrate performance on a \textit{one-to-one domain} set-up. Dataset consists of $1136$ frames at x40 magnification which are stained with standard H\&E dyes. Same tissue section has been scanned by two slide scanners: Aperio Scanscope XT and Hamamatsu Nanozoomer 2.0-HT. For evaluating the model, $500 \times 500$ patches were generated from whole slide images of both scanners resized to $256 \times 256$. $23000$ patches were used for training the model and $9600$ patches were used for evaluation. It was ensured distinct WSIs were used for training and validation to avoid any type of bias and leak.

\subsection{Network Architecture}
The generator network is a modified \textbf{U-Net} \cite{unet} which has been shown to generate excellent results in image translation tasks \cite{pix2pix}. U-Net is encoder-decoder network \cite{encdec} that uses skip connections between layers $i$ and $n-i$ where $n$ is the total number of layers in the network. In previous encoder-decoder architectures \cite{pathak2016context,wang2016generative,yoo2016pixel}. The input is passed through a series of convolutional layers that downsample the input until a bottleneck is reached after which the information is upsampled to generate an output of the desired dimensions. Therefore, by design all information passes through the bottleneck. In the stain normalization task, input and output of the network share a lot of general information that might get obscured through the flow of such a network. Skip connections in a U-Net solve this problem by circumventing the bottleneck and concatenating the output from the encoder layers to the input of the corresponding decoder layers.

The discriminator is a $4$ block CNN, which eventually outputs the decision for each image. Every convolutional block in both the generator and the discriminator is a module consisting of a convolution-normalization-ReLU layers in that order. Both instance \cite{instance} and batch \cite{batch} normalization were used; and batch normalization was empirically chosen for the final network. The convolutional layers have a kernel size of $4$ and stride $2$, with the exception of the last layer in the discriminator which operates with stride $1$.

\textbf{Self-attention layers} \cite{parikh2016decomposable} were added after every convolutional block in both the generator and the discriminator network. The self-attention mechanism complements the convolutions by establishing and leveraging long range dependencies across image regions. It help the generator synthesize images with finer details in regions based on a different spatial region in the image. Additionally the discriminator with self-attention layers is able to enforce more complex structural constraints on input images while making a decision. As described in SAGAN \cite{sagan}, a non-local network \cite{wang2018non} was used to apply the self-attention computation. The input features $x \in \mathbb{R}^{C\times N}$ are transformed using three different learnable functions $q(x), k(x), v(x)$ analogous to query, key and value setup in \cite{vaswani2017attention} as follows:
\begin{align}
    q(x) &= W_{q}x; & k(x) &= W_{k}x; &    v(x) &= W_{v}x
\end{align}
where $W_q \in \mathbb{R}^{\bar{C} \times C}$, $W_k \in \mathbb{R}^{\bar{C} \times C}$, and $W_v \in \mathbb{R}^{\bar{C} \times C}$. Also, $C$ is the number of channels, $N = height*width$ of the feature map from the previous layer and $\bar{C}$ is an adjustable parameter. For our model, $\bar{C}$ was set as $C/8$. The attention map is further calculated as:
\begin{equation}
    \begin{split}
    \alpha_{j,i} &= softmax(k(x_i)^T g(x_j)) \\ 
     &= \frac{\exp{(k(x_i)^T g(x_j))}}{\sum_{i=1}^N \exp{(k(x_i)^T g(x_j))}}    
    \end{split}
\end{equation}
where $\alpha_{j,i}$ represents the attention placed on location $i$ while synthesizing location $j$. The ouput $o \in \mathbb{R}^{C \times N}$ is calculated as:
\begin{equation}
    o_j = \sum_{i=1}^{N}\alpha_{j,i}v(x_i)
\end{equation}
The output $o$ is then scaled and added to the initial input to give the final result, 
\begin{equation}
    y_i = \mu o_i + x_i
\end{equation}
where $\mu$ is a learnable parameter that is initialized to 0.

\textbf{Spectral normalization} when applied on the layers of the discriminator network has been shown to stabilize the training of a GAN \cite{miyato2018spectral}. Moreover, based on the findings about the effect of a generator's conditioning on its performance, Zhang \cite{sagan} argue that while training a self-attention based GAN, both the generator and the discriminator can benefit from using spectral normalization. Therefore, a spectral normalization (with spectral norm of all weight layers as $1$) was added to all the networks.

\subsection{Training Details}
The parameter values of $\alpha = 10$, $\beta = 10$, $\gamma = 10$ and $\delta = 0.1$ were empirically chosen after experimentation for the evaluation model. Across all experiments, we used the Adam optimizer \cite{adam} with a learning rate of $0.0002$ and batch size $16$. The model was trained for the first $50$ epochs with a fixed learning rate and the next $50$ epochs while linearly decaying the learning rate to $0$. Instead of updating the discriminator with an image generated form the latest generator, a random image selected from a buffer of $50$ previously generated images was used to perform the update cycle \cite{shrivastava2017learning}. Least-squares adversarial loss inspired from LSGAN \cite{lsgan} was used instead of the described cross-entropy loss for some experiments. The least-squares loss stabilized the training but there was no significant visual difference in the results produced.

\begin{figure*}[t]
\centering
\includegraphics[width=0.9\textwidth]{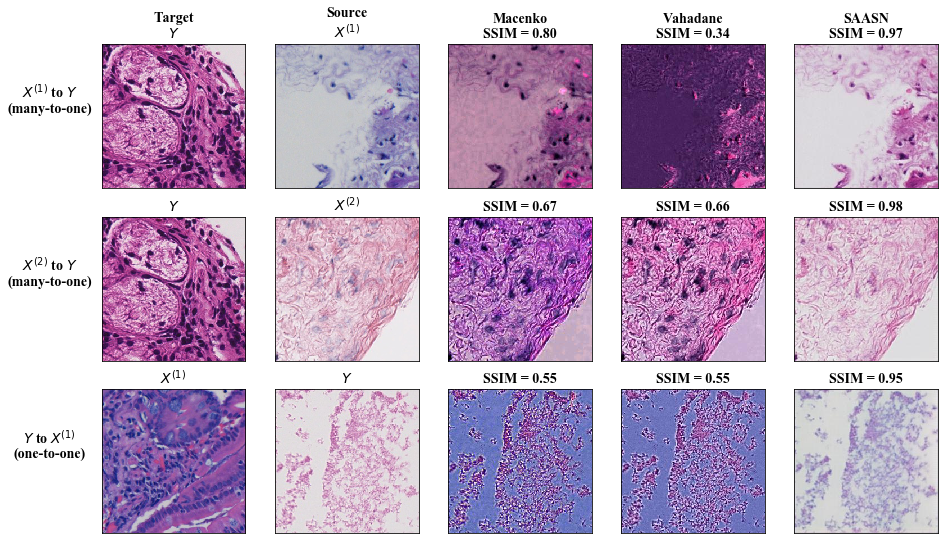} 
\caption{Visual comparison of performance in cases where Macenko and Vahadane techniques struggle to properly transfer stain in each scenario. The target image only applies to the Macenko and Vahadane techniques.}
\label{worst_viz}
\end{figure*}

\section{Results and Evaluation}

\begin{figure*}[t]
\centering
\includegraphics[width=\textwidth]{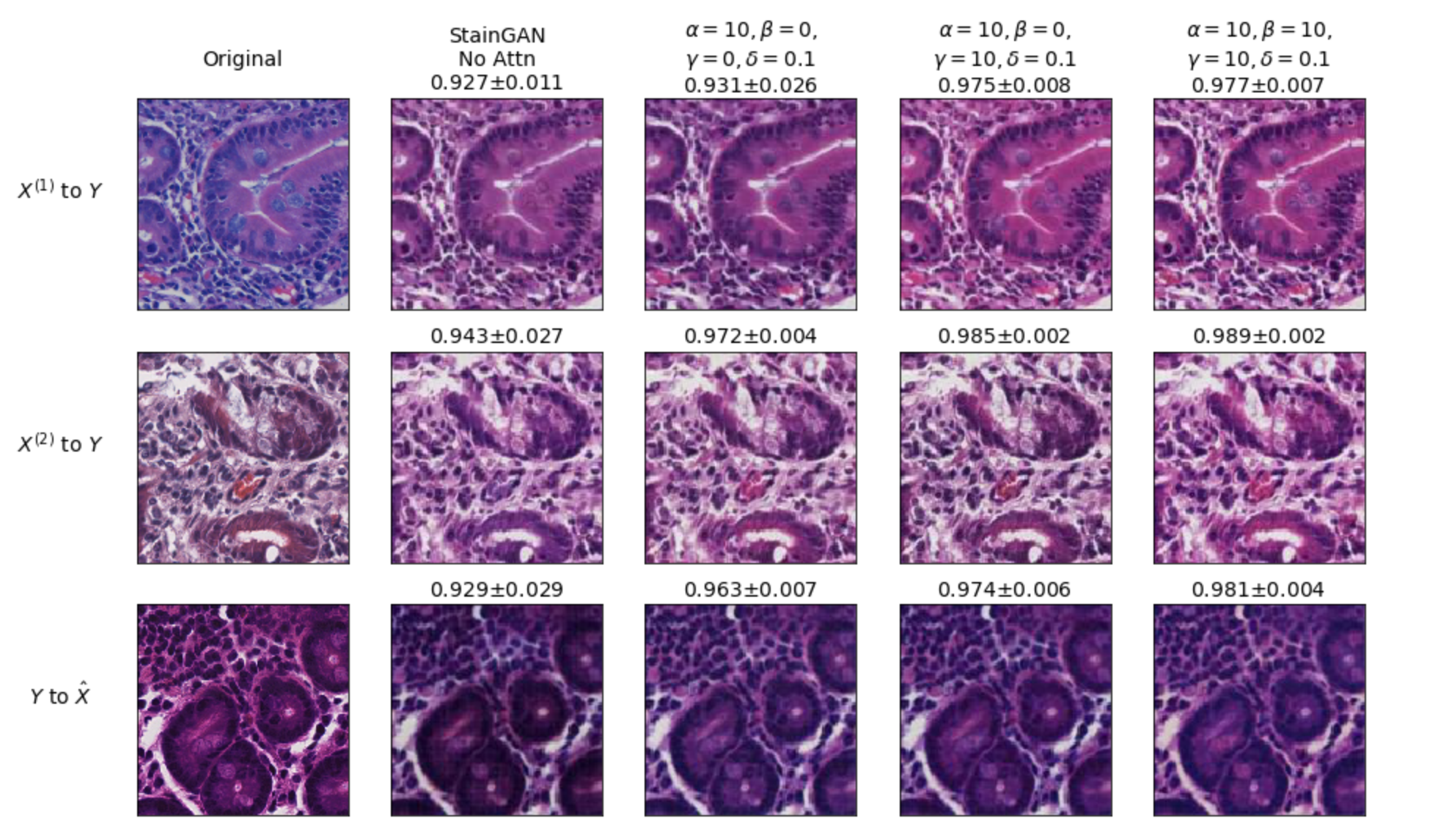} 
\caption{Visual and quantitative comparison of performance between StainGAN and ablation study on SAASN. The numbers indicate the overall mean $\pm$ standard deviation of the SSIM index for the transformation. All models were trained in a \textit{many-to-one} setup.}
\label{gan_comp}
\end{figure*}

To demonstrate the value of each introduced term in the designed loss function, an ablation study was performed. A competitive version of StainGAN \cite{shaban2019staingan} was also implemented based on the information given in the paper. It was observed that the addition of self-attention layers helped the model to generate more vibrant results that preserved medically significant artifacts. For instance, the red blood cells in the second row of Figure \ref{gan_comp} get visually merged with the surrounding cells when self-attention is not used. The ablation study shows shows that with the cycle consistency loss alone the forward mapping function ($G_{XY}$) is suppressed from providing a \textit{many-to-one} mapping as the generated domain ($\hat{X}$) from the inverse function ($G_{YX}$) will overlap more with the dominant domain in the training set. Addition of the structural cycle consistency loss term alleviates this issue as it is stain agnostic and a combination of the said losses gives a more compelling result. 

To evaluate the stain transfer, the Structural Similarity (SSIM) index is again utilized. SSIM is calculated by comparing the normalized image with the original. Both images are converted to gray-scale before beginning SSIM calculations. Our approach is compared to two of the most popular unsupervised stain normalization techniques, Macenko \cite{macenko2009method} and Vahadane \cite{vahadane2016structure}. The popular supervised approach by Khan \cite{khan2014nonlinear} could not be tested due to lack of pixel-level labeling in our data. These results are compiled in Table \ref{table1} for duodenal biopsy dataset and in Table \ref{table2} for MITOS-ATYPIA dataset. For the $X^{(1)}$ to $Y$ and the $X^{(2)}$ to $Y$ stain transfers, the values for SAASN are higher than the other two normalization techniques and the variance is significantly smaller. This demonstrates that SAASN is not only better at preserving structure, but also consistently transfers stain without major anomalies. The traditional approaches (Vahadane and Macenko) approaches can struggle if the source has a much different stain distribution than the target. This can lead to the stains appearing in the wrong areas on the normalized image. SAASN is able to leverage information from entire stain domains and therefore is not as affected by this issue. These results demonstrate that SAASN can be trusted to produce consistent stain transfers on a robust set of stain patterns in WSI patches.

\begin{table}
\scriptsize
\begin{center}
\begin{tabular}{|l|c|c|c|}
\hline
Method & $X^{(1)}$ to $Y$ & $X^{(2)}$ to $Y$ & $Y$ to $X^{(1)}$\\
\hline
Vahadane & $0.861 \pm 0.108$ & $0.919 \pm 0.029$ & $0.932 \pm 0.033$\\
Macenko & $0.942 \pm 0.033$ & $0.934 \pm 0.022$ & $0.941 \pm 0.020$\\
StainGAN & $0.927 \pm 0.011$ & $0.943 \pm 0.027$ & $0.929 \pm 0.021$\\
SAASN  & \bm{$0.977 \pm 0.007$} & \bm{$0.989 \pm 0.002$} & \bm{$0.981 \pm 0.004$}\\
\hline
\end{tabular}
\end{center}
\caption{Mean $\pm$ Standard deviation of the SSIM index values for normalization across domains. For StainGAN and SAASN all values are computed for a \textit{many-to-one} setup on the first dataset.}
\label{table1}
\end{table}

\begin{figure*}[t]
\centering
\includegraphics[width=0.90\textwidth]{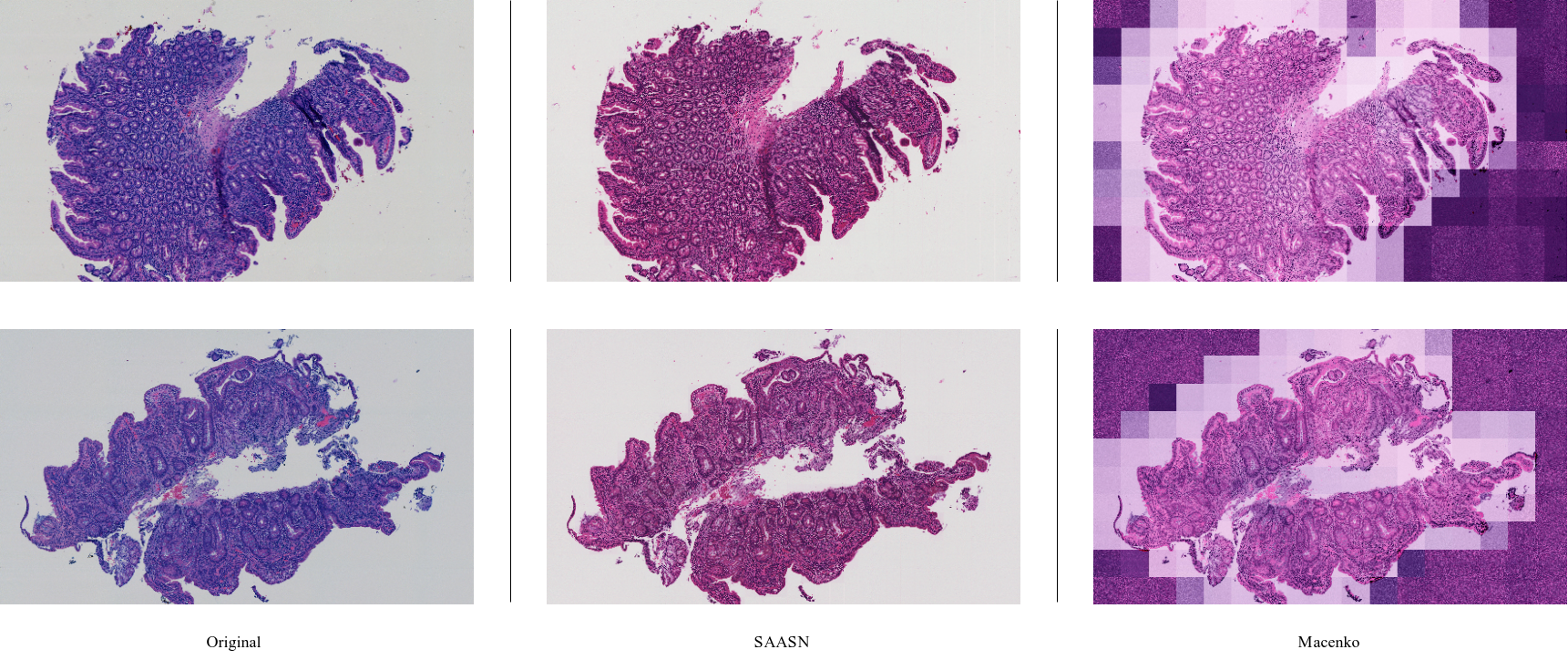} 
\caption{Normalized Whole Slide Image using ours and traditional approaches. Macenko was chosen because it performed better than Vahadane on our dataset. The target slide for Macenko was empirically selected to give the best translation.}
\label{trad_comp}
\end{figure*}

In addition to assessing the structure-preserving ability of the stain normalization methods, visual comparisons are essential to ensure that the stains have transferred properly. In Figure \ref{worst_viz}, results are displayed for the three stain transfers in duodenal biopsy dataset. The images with the smallest $L2$-norm for combined Macenko and Vahadane SSIM values were selected to demonstrate the performance of SAASN. For $X^{(1)}$ to $Y$ and $X^{(2)}$ to $Y$, the same target image from domain $Y$ is used. For $Y$ to $X^{(1)}$, a target image from domain $X^{(1)}$ is used. The three selected source images are similar in that they all have a large majority of pixels containing connective tissue or background. The unsupervised approaches can struggle executing color deconvolution on these types of images. This is apparent in the Macenko and Vahadane normalizations shown in Figure \ref{worst_viz}. The stains are either inverted (hematoxylin-like color transferred to the background) or confusing connective tissue as an actual cell structure. Meanwhile, SAASN did not have difficulty identifying the connective tissue or background pixels in the source image. 

\begin{table}
\scriptsize
\begin{center}
\begin{tabular}{|l|c|c|c|}
\hline
Method & Aperio to Hamamatsu & Hamamatsu to Aperio\\
\hline
Vahadane & $0.971 \pm 0.031$ & $0.955 \pm 0.038$\\
Macenko & $0.968 \pm 0.034$ & $0.956 \pm 0.039$\\
StainGAN & $0.967 \pm 0.009$ & $0.947 \pm 0.032$\\
SAASN  & \bm{$0.995 \pm 0.001$} & \bm{$0.996 \pm 0.001$}\\
\hline
\end{tabular}
\end{center}
\caption{Mean $\pm$ Standard deviation of the SSIM index values for normalization across domains. For StainGAN and SAASN all values are computed for the \textit{one-to-one} setup on the second dataset.}
\label{table2}
\end{table}

A similar analysis was also performed using the highest $L2$-norm values. These are the examples where the traditional methods performed the best\footnote{Please refer to \href{https://github.com/4m4n5/saasn-stain-normalization}{https://github.com/4m4n5/saasn-stain-normalization} and additional materials for implementation and additional results.}. Vahadane and Macenko are able to maintain structure, but may not visually match the target distribution or the proper background pixel color.

Stain normalization is crucial for bias-free visual examination of Whole Slide Images (WSIs) and diagnosis by medical practitioners in control trial settings. WSIs have very large dimensions and cannot be normalized without resizing to a computationally tractable size which results in a significant loss in resolution. To normalize WSIs, they must be split into patches, normalized and then stitched back together. Traditional methods perform computations for transformation independently on these patches. As a result, it is impossible to reconstruct a WSI that has a consistent stain and is indistinguishable from an original image in the target domain. As demonstrated in Figure \ref{trad_comp}, for our method, since the trained weights of the mapping function are constant during this transformation, the reconstructed WSI could not be distinguished from original images and thus is easier for medical professionals to hold diagnosis trails.

In order to a validate a successful translation three medical professionals, including a board-certified pathologist, completed a blind review of 10 WSIs normalized via traditional and our method as shown in Figure \ref{trad_comp}. The pathologist confirmed that medically relevant cell types (polymorphonuclear neutrophils, epithelial cells, eosinophils, lymphocytes, goblet cells, paneth cells, neuroendocrine cells) were not lost during translation. The pathologist further observed that our method was able to completely preserve the structure and the density of all of these cell types which traditional methods only partially preserved. Specifically, the eosinophilic granules in paneth cells, neuroendocrine cells and eosinophils were not appreciated in traditionally stain normalized WSIs which made it difficult to differentiating these cells from each other.
cifically, the eosinophilic granules in paneth cells, neuroendocrine cells and eosinophils were not appreciated in traditionally stain normalized WSIs which made it difficult to differentiating these cells from each other.aaaaaaaaahhhh

%
%
%
{
\bibliographystyle{splncs04}
\bibliography{mybibliography}

\begin{thebibliography}{10}
\providecommand{\url}[1]{\texttt{#1}}
\providecommand{\urlprefix}{URL }
\providecommand{\doi}[1]{https://doi.org/#1}

\bibitem{bejnordi2014quantitative}
Bejnordi, B.E., Timofeeva, N., Otte-H{\"o}ller, I., Karssemeijer, N., van~der
  Laak, J.A.: Quantitative analysis of stain variability in histology slides
  and an algorithm for standardization. In: Medical Imaging 2014: Digital
  Pathology. vol.~9041, p. 904108. International Society for Optics and
  Photonics (2014)

\bibitem{bentaieb2017adversarial}
BenTaieb, A., Hamarneh, G.: Adversarial stain transfer for histopathology image
  analysis. IEEE transactions on medical imaging  \textbf{37}(3),  792--802
  (2017)

\bibitem{coltuc2006exact}
Coltuc, D., Bolon, P., Chassery, J.M.: Exact histogram specification. IEEE
  Transactions on Image Processing  \textbf{15}(5),  1143--1152 (2006)

\bibitem{fischer2008hematoxylin}
Fischer, A.H., Jacobson, K.A., Rose, J., Zeller, R.: Hematoxylin and eosin
  staining of tissue and cell sections. Cold Spring Harbor Protocols
  \textbf{2008}(5),  pdb--prot4986 (2008)

\bibitem{gan}
Goodfellow, I., Pouget-Abadie, J., Mirza, M., Xu, B., Warde-Farley, D., Ozair,
  S., Courville, A., Bengio, Y.: Generative adversarial nets. In: Advances in
  neural information processing systems. pp. 2672--2680 (2014)

\bibitem{encdec}
Hinton, G.E., Salakhutdinov, R.R.: Reducing the dimensionality of data with
  neural networks. science  \textbf{313}(5786),  504--507 (2006)

\bibitem{batch}
Ioffe, S., Szegedy, C.: Batch normalization: Accelerating deep network training
  by reducing internal covariate shift. arXiv preprint arXiv:1502.03167  (2015)

\bibitem{pix2pix}
Isola, P., Zhu, J.Y., Zhou, T., Efros, A.A.: Image-to-image translation with
  conditional adversarial networks. In: Proceedings of the IEEE conference on
  computer vision and pattern recognition. pp. 1125--1134 (2017)

\bibitem{janowczyk2017stain}
Janowczyk, A., Basavanhally, A., Madabhushi, A.: Stain normalization using
  sparse autoencoders (stanosa): Application to digital pathology. Computerized
  Medical Imaging and Graphics  \textbf{57},  50--61 (2017)

\bibitem{khan2014nonlinear}
Khan, A.M., Rajpoot, N., Treanor, D., Magee, D.: A nonlinear mapping approach
  to stain normalization in digital histopathology images using image-specific
  color deconvolution. IEEE Transactions on Biomedical Engineering
  \textbf{61}(6),  1729--1738 (2014)

\bibitem{kim2017learning}
Kim, T., Cha, M., Kim, H., Lee, J.K., Kim, J.: Learning to discover
  cross-domain relations with generative adversarial networks. In: Proceedings
  of the 34th International Conference on Machine Learning-Volume 70. pp.
  1857--1865. JMLR. org (2017)

\bibitem{adam}
Kingma, D.P., Ba, J.: Adam: A method for stochastic optimization. arXiv
  preprint arXiv:1412.6980  (2014)

\bibitem{litjens2017survey}
Litjens, G., Kooi, T., Bejnordi, B.E., Setio, A.A.A., Ciompi, F., Ghafoorian,
  M., Van Der~Laak, J.A., Van~Ginneken, B., S{\'a}nchez, C.I.: A survey on deep
  learning in medical image analysis. Medical image analysis  \textbf{42},
  60--88 (2017)

\bibitem{liu2017unsupervised}
Liu, M.Y., Breuel, T., Kautz, J.: Unsupervised image-to-image translation
  networks. In: Advances in neural information processing systems. pp. 700--708
  (2017)

\bibitem{45922}
Liu, Y., Gadepalli, K.K., Norouzi, M., Dahl, G., Kohlberger, T., Venugopalan,
  S., Boyko, A.S., Timofeev, A., Nelson, P.Q., Corrado, G., Hipp, J., Peng, L.,
  Stumpe, M.: Detecting cancer metastases on gigapixel pathology images. Tech.
  rep., arXiv (2017), \url{https://arxiv.org/abs/1703.02442}

\bibitem{macenko2009method}
Macenko, M., Niethammer, M., Marron, J.S., Borland, D., Woosley, J.T., Guan,
  X., Schmitt, C., Thomas, N.E.: A method for normalizing histology slides for
  quantitative analysis. In: 2009 IEEE International Symposium on Biomedical
  Imaging: From Nano to Macro. pp. 1107--1110. IEEE (2009)

\bibitem{lsgan}
Mao, X., Li, Q., Xie, H., Lau, R.Y., Wang, Z., Paul~Smolley, S.: Least squares
  generative adversarial networks. In: Proceedings of the IEEE International
  Conference on Computer Vision. pp. 2794--2802 (2017)

\bibitem{miyato2018spectral}
Miyato, T., Kataoka, T., Koyama, M., Yoshida, Y.: Spectral normalization for
  generative adversarial networks. arXiv preprint arXiv:1802.05957  (2018)

\bibitem{parikh2016decomposable}
Parikh, A.P., T{\"a}ckstr{\"o}m, O., Das, D., Uszkoreit, J.: A decomposable
  attention model for natural language inference. arXiv preprint
  arXiv:1606.01933  (2016)

\bibitem{pathak2016context}
Pathak, D., Krahenbuhl, P., Donahue, J., Darrell, T., Efros, A.A.: Context
  encoders: Feature learning by inpainting. In: Proceedings of the IEEE
  conference on computer vision and pattern recognition. pp. 2536--2544 (2016)

\bibitem{reinhard2001color}
Reinhard, E., Adhikhmin, M., Gooch, B., Shirley, P.: Color transfer between
  images. IEEE Computer graphics and applications  \textbf{21}(5),  34--41
  (2001)

\bibitem{unet}
Ronneberger, O., Fischer, P., Brox, T.: U-net: Convolutional networks for
  biomedical image segmentation. In: International Conference on Medical image
  computing and computer-assisted intervention. pp. 234--241. Springer (2015)

\bibitem{roy2018study}
Roy, S., kumar Jain, A., Lal, S., Kini, J.: A study about color normalization
  methods for histopathology images. Micron  \textbf{114},  42--61 (2018)

\bibitem{shaban2019staingan}
Shaban, M.T., Baur, C., Navab, N., Albarqouni, S.: Staingan: Stain style
  transfer for digital histological images. In: 2019 IEEE 16th International
  Symposium on Biomedical Imaging (ISBI 2019). pp. 953--956. IEEE (2019)

\bibitem{shrivastava2017learning}
Shrivastava, A., Pfister, T., Tuzel, O., Susskind, J., Wang, W., Webb, R.:
  Learning from simulated and unsupervised images through adversarial training.
  In: Proceedings of the IEEE conference on computer vision and pattern
  recognition. pp. 2107--2116 (2017)

\bibitem{identity}
Taigman, Y., Polyak, A., Wolf, L.: Unsupervised cross-domain image generation.
  arXiv preprint arXiv:1611.02200  (2016)

\bibitem{instance}
Ulyanov, D., Vedaldi, A., Lempitsky, V.: Instance normalization: The missing
  ingredient for fast stylization. arXiv preprint arXiv:1607.08022  (2016)

\bibitem{vahadane2016structure}
Vahadane, A., Peng, T., Sethi, A., Albarqouni, S., Wang, L., Baust, M.,
  Steiger, K., Schlitter, A.M., Esposito, I., Navab, N.: Structure-preserving
  color normalization and sparse stain separation for histological images. IEEE
  transactions on medical imaging  \textbf{35}(8),  1962--1971 (2016)

\bibitem{vaswani2017attention}
Vaswani, A., Shazeer, N., Parmar, N., Uszkoreit, J., Jones, L., Gomez, A.N.,
  Kaiser, {\L}., Polosukhin, I.: Attention is all you need. In: Advances in
  neural information processing systems. pp. 5998--6008 (2017)

\bibitem{wang2018non}
Wang, X., Girshick, R., Gupta, A., He, K.: Non-local neural networks. In:
  Proceedings of the IEEE Conference on Computer Vision and Pattern
  Recognition. pp. 7794--7803 (2018)

\bibitem{wang2016generative}
Wang, X., Gupta, A.: Generative image modeling using style and structure
  adversarial networks. In: European Conference on Computer Vision. pp.
  318--335. Springer (2016)

\bibitem{ssim}
Wang, Z., Bovik, A.C., Sheikh, H.R., Simoncelli, E.P., et~al.: Image quality
  assessment: from error visibility to structural similarity. IEEE transactions
  on image processing  \textbf{13}(4),  600--612 (2004)

\bibitem{wei2019automated}
Wei, J.W., Wei, J.W., Jackson, C.R., Ren, B., Suriawinata, A.A., Hassanpour,
  S.: Automated detection of celiac disease on duodenal biopsy slides: A deep
  learning approach. Journal of pathology informatics  \textbf{10} (2019)

\bibitem{yi2017dualgan}
Yi, Z., Zhang, H., Tan, P., Gong, M.: Dualgan: Unsupervised dual learning for
  image-to-image translation. In: Proceedings of the IEEE international
  conference on computer vision. pp. 2849--2857 (2017)

\bibitem{yoo2016pixel}
Yoo, D., Kim, N., Park, S., Paek, A.S., Kweon, I.S.: Pixel-level domain
  transfer. In: European Conference on Computer Vision. pp. 517--532. Springer
  (2016)

\bibitem{zanjani2018stain}
Zanjani, F.G., Zinger, S., Bejnordi, B.E., van~der Laak, J.A., de~With, P.H.:
  Stain normalization of histopathology images using generative adversarial
  networks. In: 2018 IEEE 15th International Symposium on Biomedical Imaging
  (ISBI 2018). pp. 573--577. IEEE (2018)

\bibitem{sagan}
Zhang, H., Goodfellow, I., Metaxas, D., Odena, A.: Self-attention generative
  adversarial networks. arXiv preprint arXiv:1805.08318  (2018)

\bibitem{cyclegan}
Zhu, J.Y., Park, T., Isola, P., Efros, A.A.: Unpaired image-to-image
  translation using cycle-consistent adversarial networks. In: Proceedings of
  the IEEE international conference on computer vision. pp. 2223--2232 (2017)

\end{thebibliography}
}
\appendix

\section{Conclusions}
The proposed framework is successful in effective translation of the stain appearance of histopathological images while preserving the biological features in the process. This setup was specifically designed to accommodate a \textit{many-to-one} stain transfer situation in which multiple stains are converted to a common domain. SAASN is compared to other leading stain normalization techniques on pathology images in both \textit{one-to-one} and \textit{many-to-one} setup. SAASN consistently performed successful stain transfers even when the other techniques failed due to large variations between the source and target image stains and unconventional input image structures. Results also show that SAASN outperformed traditional methods at preserving the cellular structures. We contend that the proposed unsupervised image to image translation approach can be successfully applied to general \textit{many-to-one} image translation problems outside the medical domain as well. 

\begin{figure*}[t]
\centering
\includegraphics[width=\textwidth]{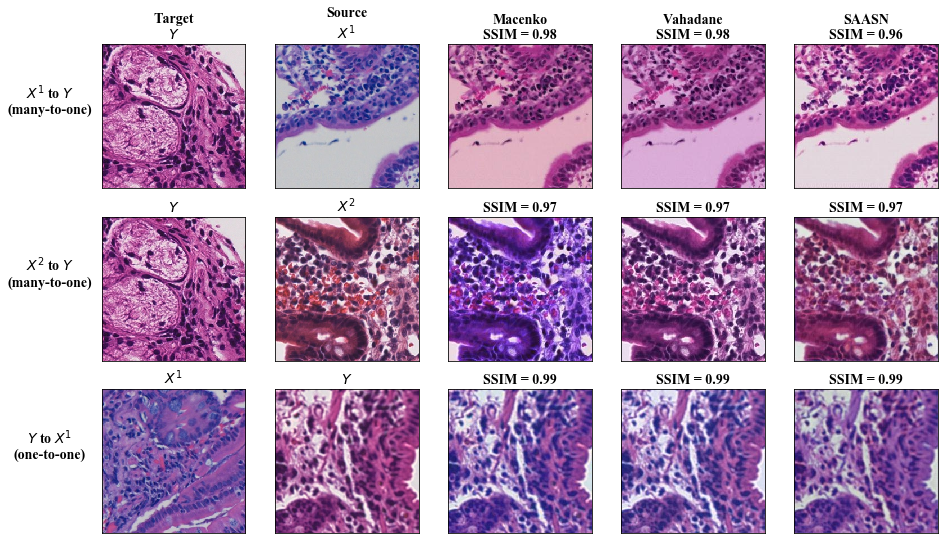} 
\caption{Visual comparison of performance in cases where Macenko and Vahadane techniques perform very well according to a combined SSIM index. The target image only applies to the Macenko and Vahadane techniques. The main results section included a visual comparison of SAASN stain transfers with the worst performing Macenko and Vahadane images based on SSIM. Alternatively, SAASN is also compared to the best SSIM indexes for the other two techniques. Figure \ref{viz_best} displayed the top three images in each stain transfer scenario based on the highest $L2$-norm of Macenko and Vahadane SSIM results. For the $X^{(1)}$ to $Y$ transfer, SAASN was the only technique that properly maintained a whitish/gray background pixel color. For the $X^{(2)}$ to $Y$ transfer, Macenko appeared to create a new stain distribution that was not close to the desired target image. All three normalizations performed well in the \textit{one-to-one} transfer. The comparison in Figure \ref{viz_best} demonstrates that SAASN can perform better at preserving structure and properly transferring stain domains, because both areas are incorporated into the network's loss functions.}
\label{viz_best}
\end{figure*}

\section{Additional results}
We trained and tested the model in both a \textit{one-to-one} ($K = 1$) and \textit{many-to-one} ($K = 2$) setup. In this section we demonstrate the model performance, on test datasets, for visual inspection.

\begin{figure*}[t]
\centering
\includegraphics[width=\textwidth]{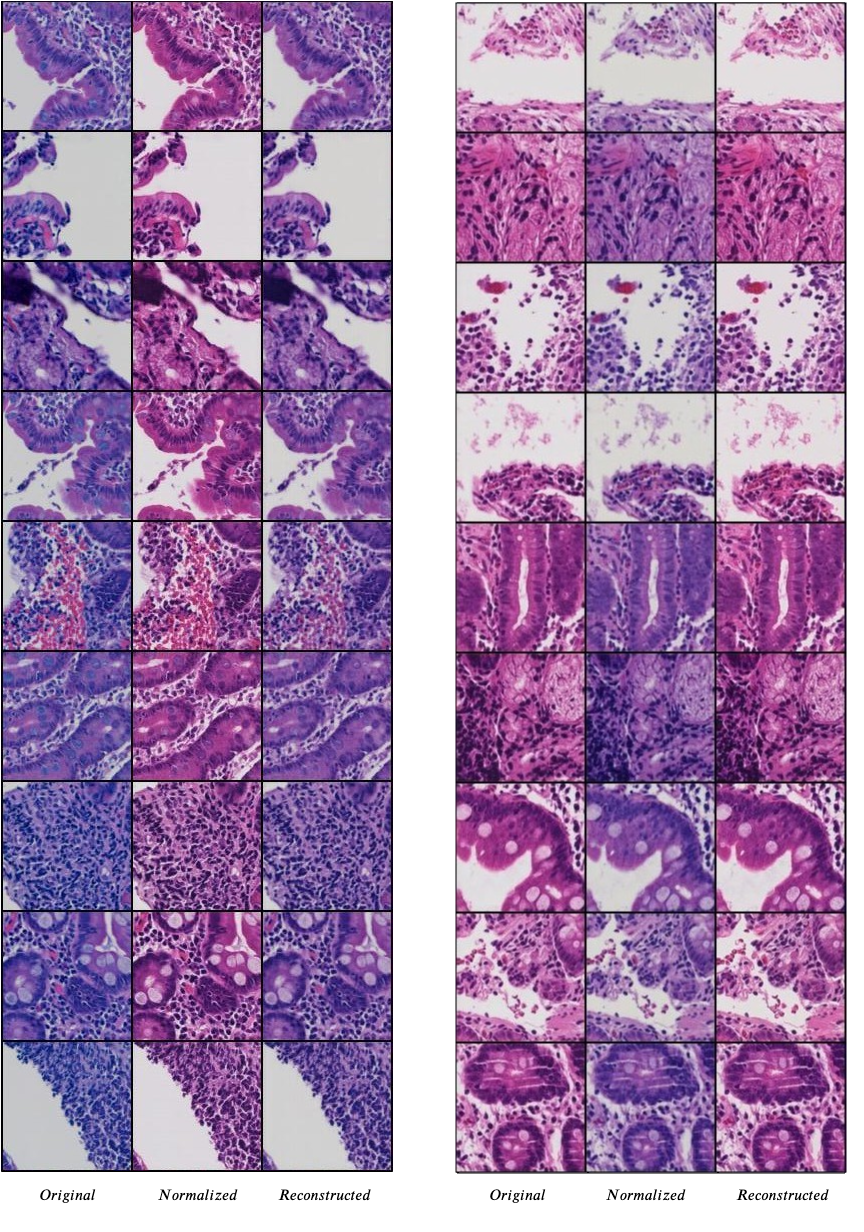} 
\caption{\textit{One-to-one} (K=1) model. \textbf{Left:} Translation from domain $X^{(1)}$ to $Y$ and back to domain $X^{(1)}$. \textbf{Right:} Translation from domain $Y$ to $X^{(1)}$ and back to $Y$.}
\label{x1_y}
\end{figure*}

\begin{figure*}[t]
\centering
\includegraphics[width=\textwidth]{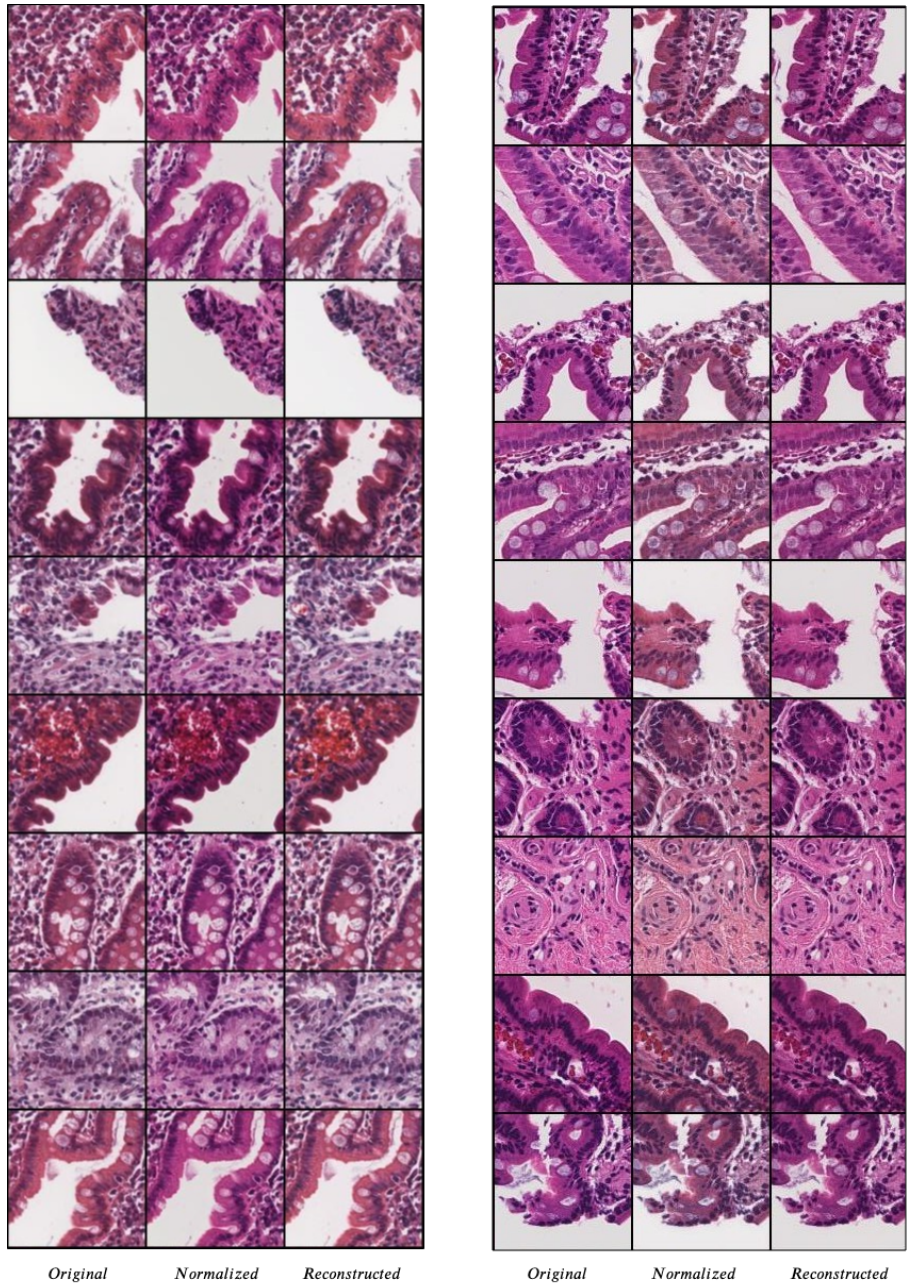} 
\caption{\textit{One-to-one} (K=1) model. \textbf{Left:} Translation from domain $X^{(2)}$ to $Y$ and back to domain $X^{(2)}$. \textbf{Right:} Translation from domain $Y$ to $X^{(2)}$ and back to $Y$.}
\label{x2_y}
\end{figure*}

\begin{figure*}[t]
\centering
\includegraphics[width=\textwidth]{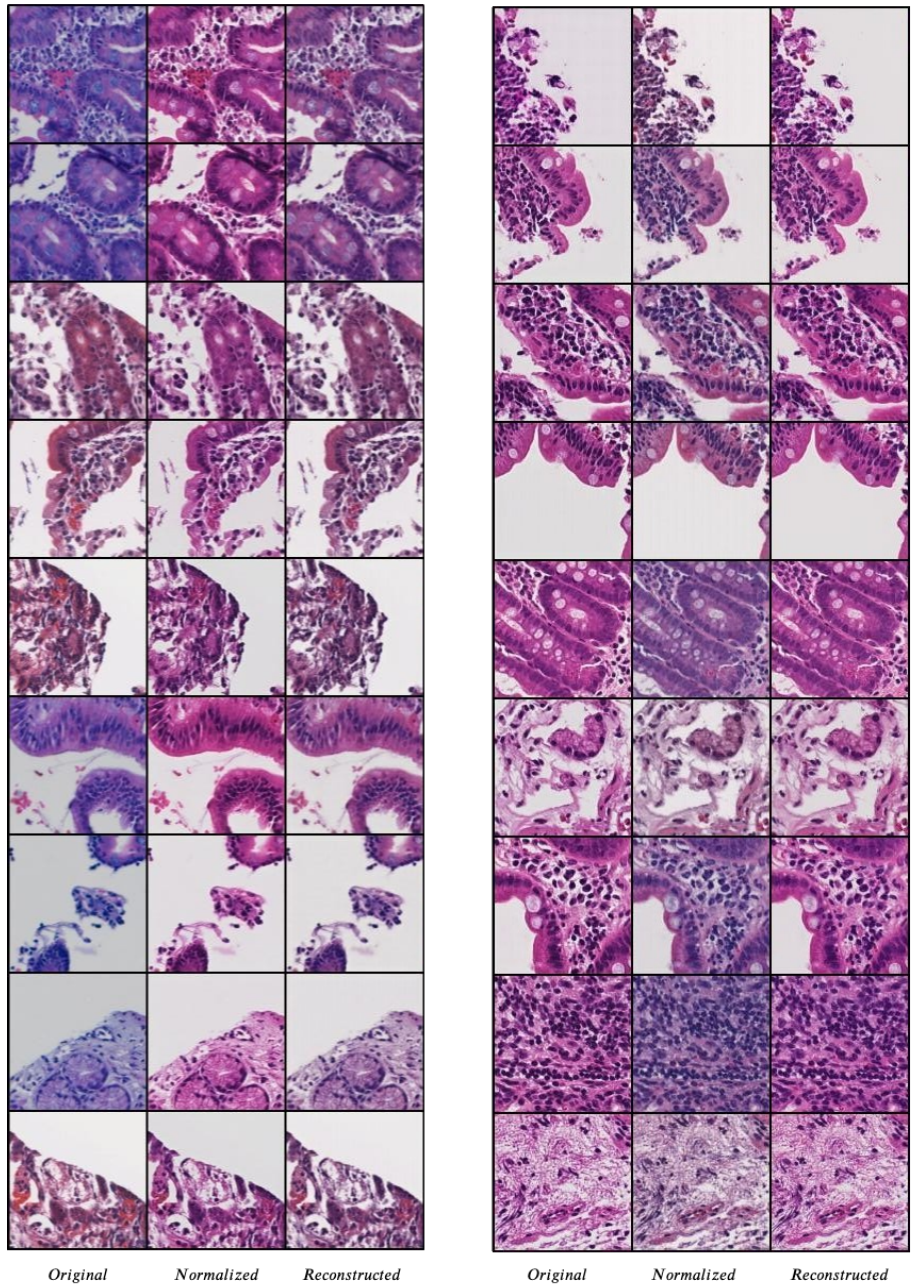} 
\caption{\textit{Many-to-one} (K=2) model. \textbf{Left:} Translation from domain $X$ to $Y$ and back to domain $\hat{X}$. \textbf{Right:} Translation from domain $Y$ to $\hat{X}$ and back to $Y$.}
\label{many_to_one}
\end{figure*}

\begin{figure*}[t]
\centering
\includegraphics[width=\textwidth]{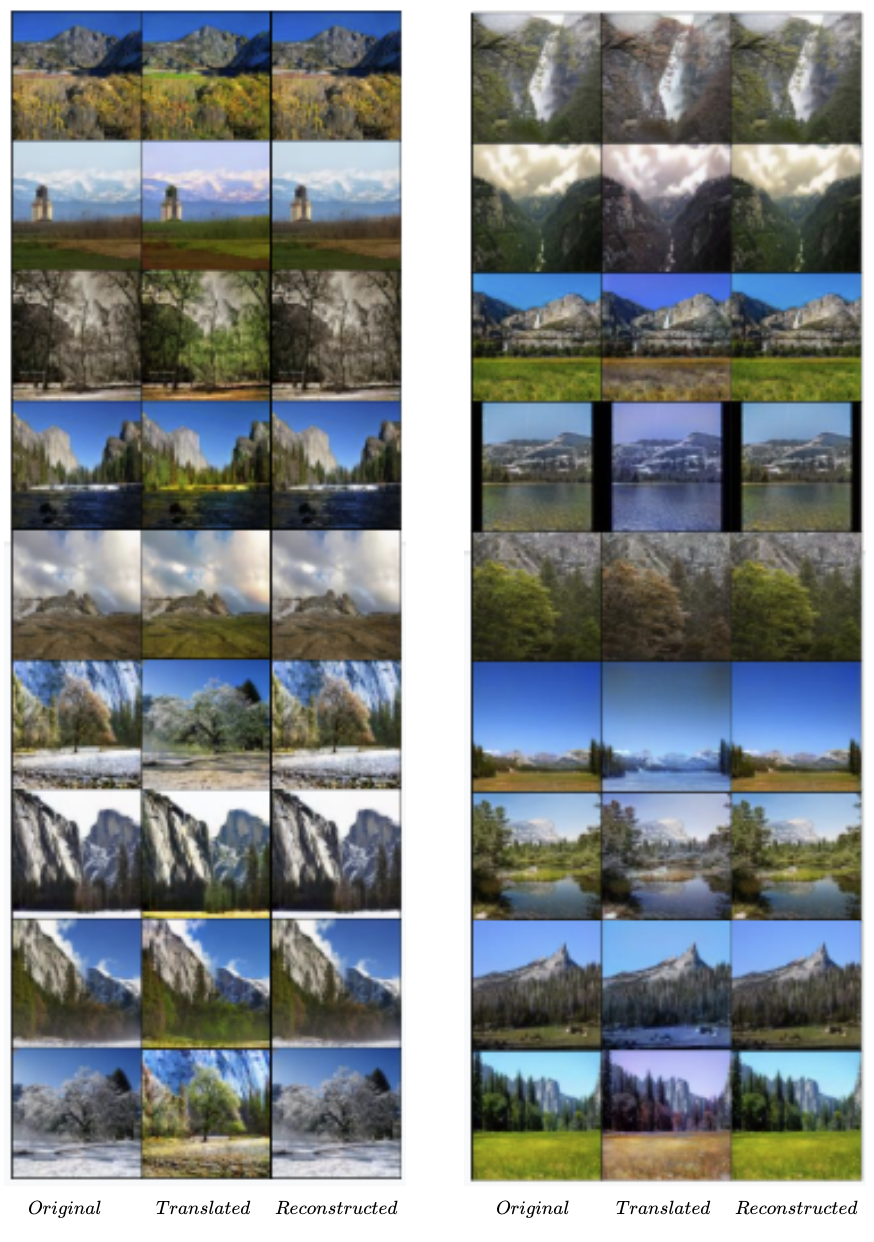} 
\caption{The model was also trained on Yosemite summer to winter dataset from the CycleGAN paper. \textbf{Left:} Translation from winter to summer and back to winter. \textbf{Right:} Translation from summer to winter and back to summer.  The model was trained with the same parameters as for the stain normalization task.}
\label{many_to_one}
\end{figure*}

\end{document}